\documentclass{iaus}
\usepackage{graphicx}

\title[Nebula around R Corona Borealis]
{Nebula around R Corona Borealis}

\author[N. Kameswara Rao \& David L. Lambert] 
{N. Kameswara Rao$^{1,2}$
 \and David L. Lambert$^1$}

\affiliation{$^1$The W. J. McDonald Observatory,The University of Texas, Austin, USA
\\
$^2$Instituto de Astrofísica de Canarias, IAC. La Laguna, Tenerife, Spain
}

\pubyear{2011}
\volume{283}  
\pagerange{119--126}
\jname{Planetary Nebulae: an Eye to the Future
}
\editors{}
\begin{document}

\maketitle

\begin{abstract}
The star R Corona Borealis (R CrB) shows forbidden lines of [O II], [N II], and [S II] during the deep minimum 
when the star is fainter by about 8 to 9 magnitudes from normal brightness, suggesting the presence of nebular 
material around it. We present low and high spectral resolution observations of these lines during the ongoing 
deep minimum of R CrB, which started in July 2007. These emission lines show double peaks with a separation of 
about 170 km/s. The line ratios of [S II] and [O II] suggest an electron density of about 100 cm$^{-3}$. We discuss 
the physical conditions and possible origins of this low density gas. These forbidden lines have also been seen 
in other R Coronae Borealis stars during their deep light minima and this is a general characteristic of these 
stars, which might have some relevance to their origins.
\end{abstract}

\firstsection 
\section{Introduction}

The R Coronae Borealis (RCB hereafter) stars are notable for two distinct characteristics: 
(1) they are hydrogen-poor (by a factor ranging from $\sim$10 to 10$^8$), helium-rich, carbon-rich, F-G(K) supergiants; 
(2) they undergo large declines in light (2 to 8 magnitudes) at irregular intervals due to formation of clouds 
of carbon dust in the line of sight to the star. 

During the deep light minimum of RCBs, the star is almost completely obscured (by dust) and the environment 
of the star is revealed. 
The spectrum at minimum light shows  
a set of broad emission lines, among other things, (FWHM$\sim$150-200 km/s)  that emerge when the star becomes faint (5 magnitudes or more), consisting 
of He I, Ca II, Na I D lines, and forbidden lines of [N II], [O II], [S II], and [Ca II]. We report on a 
study of nebular lines during the current ongoing minimum of R CrB that started in 2007 July.


\section{Present observations: 2007 -- 2011 minimum of R Cr}

Spectroscopic observations have been obtained at various resolutions using HFOSC 2m Chandra telescope at 
Hanle, Ladhak (R=2,300), 2.7m Harlan Smith telescope at McDonald Observatory (R=30-60,000) and 10-m 
Keck1 telescope (R=30,000). Forbidden lines seen are: [O II], [N II], [Ca II], and [S II]. 
They show  double peaks at high resolution and are separated by 170 km/s. The FWHM of each peak 
is 140 km/s.

Electron density and temperature (nebular lines): [S II]$\lambda$6717 and $\lambda$6731 are present in the HCT spectra 
obtained in 2009 Feb 12 (Fig.1). The 6717/6731 flux ratio of 1.4 suggests $N_e$ of about 
50-100 cm$^{-3}$ . The Keck high-resolution spectrum obtained on 2010 June 20 
covers the region of the [O II] lines (Figure 2). Since the lines are double peaked, the red and blue 
components of the [O II] $\lambda$3729 and $\lambda$3726 lines, respectively, are blended. To estimate their flux ratio, 
we assume that [O II] profiles and the flux ratio of the blue and red components are similar to that of 
the [N II] $\lambda$6583 line. We arrived at the  [O II] F(3729)/F(3726) flux ratio of 1.21 by scaling the
 [N II] $\lambda$6583 line to match the [O II] profiles that   
 suggested $Ne$ of 100 cm$^{-3}$, indicating a low density nebula. The [N II] $\lambda$6548 \& 6583 are  very close to the expected ratio (1:3) but the [N II] 
$\lambda$5755 line was never seen.  By using an upper limit to the flux of the latter line, Te $<$ 7500 K is obtained. 
The nebular extent is estimated to be about 2308 R$_{\hbox{star}}$ (0.59$^{\prime\prime}$) - by adopting a  
distance of 1.6 kpc  which is 
much bigger than the location of the hot dust ($\sim$600 K). Once ionized, at these low density, the nebula 
remains ionized, even if the source of ionization is cut off, (for about 10$^4$ years).

Variability of  [N II] during the minimum: the [N II] $\lambda$6583 flux seemed to have varied along with the 
profile during the current minimum. The flux in 1995-96 minimum was 1.96$\times$10$^{-17}$ W m$^{-2}$. The same flux is 
also seen in the initial part of the current minimum but has decreased in a two years period from 
1.97$\times$10$^{-17}$ to 1.16$\times$10$^{-17}$ W m$^{-2}$. The red peak flux decreased by 2010 June 20. 
The velocity range on 
the red side also decreased  from 200 to 170 km/s. What caused the eating of the red side of 
the profile? (This is seen in other permitted line profiles too - (Na I D, Ca II, He I, etc.) 
- movement of a dust cloud? suggests that the blue emission (expanding gas) is closer to us than new  dust?

\begin{figure}[t]
\begin{center}
 \includegraphics[width=1.7in]{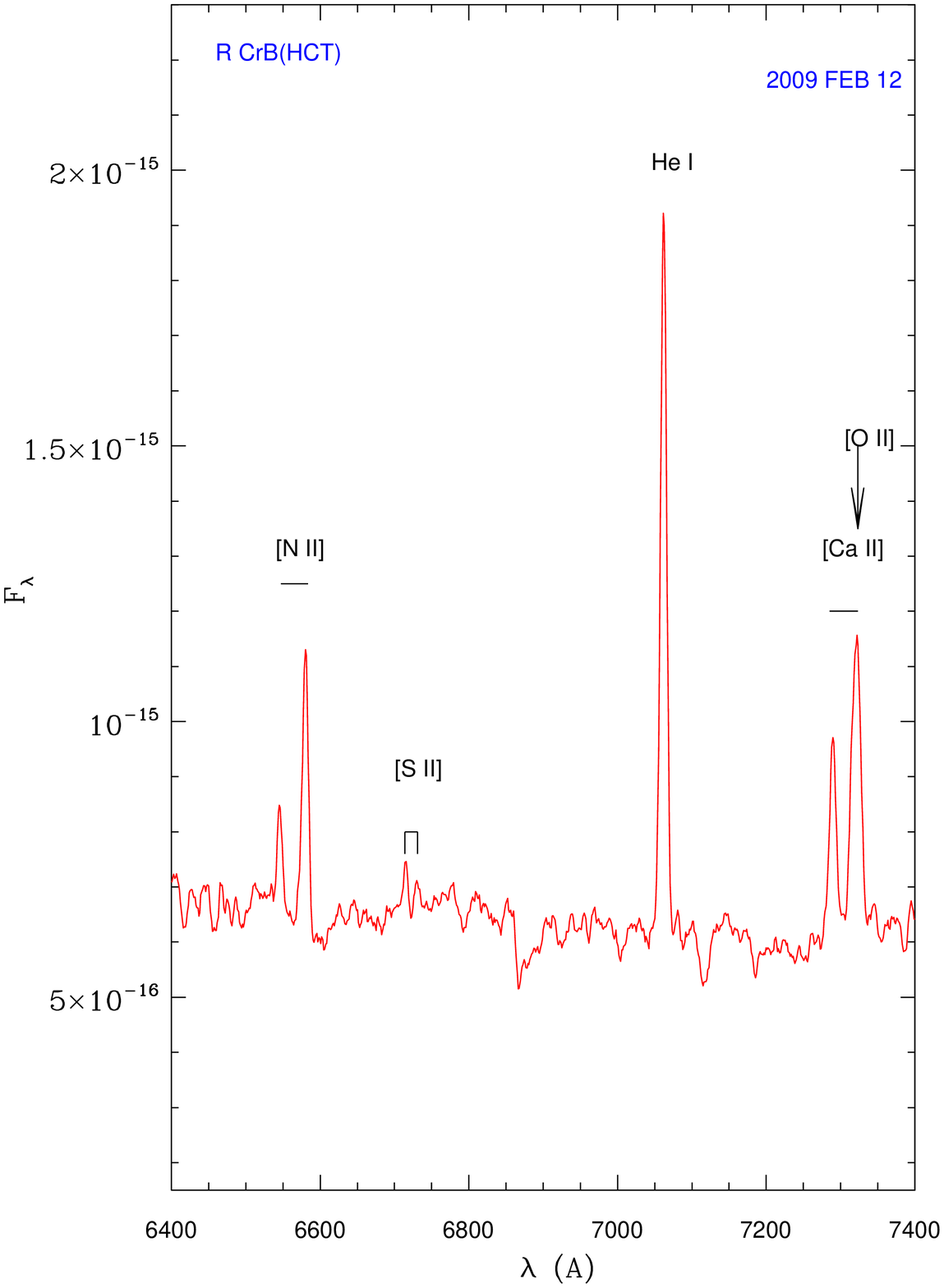}
 \includegraphics[width=1.7in]{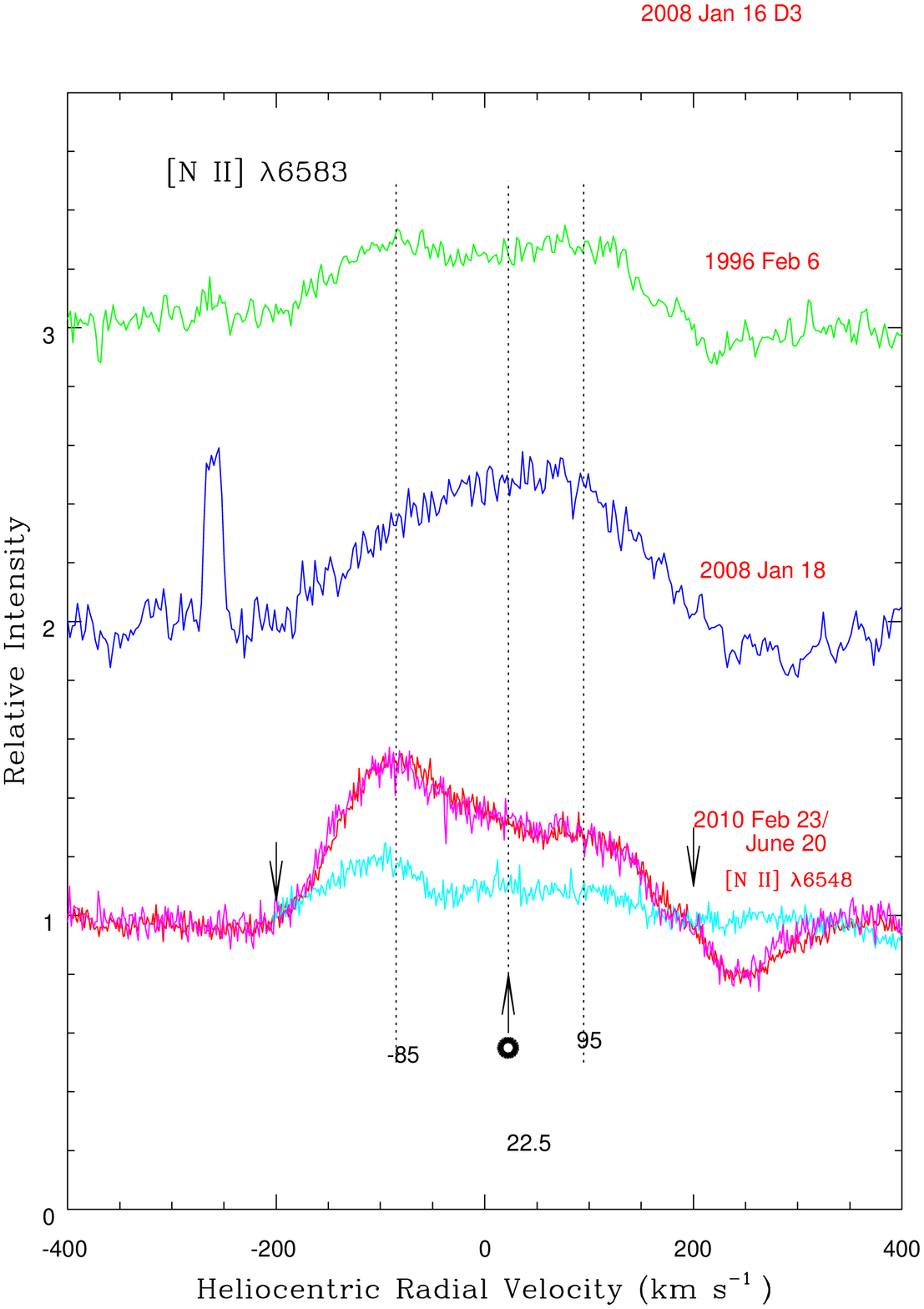}
 \includegraphics[width=1.7in]{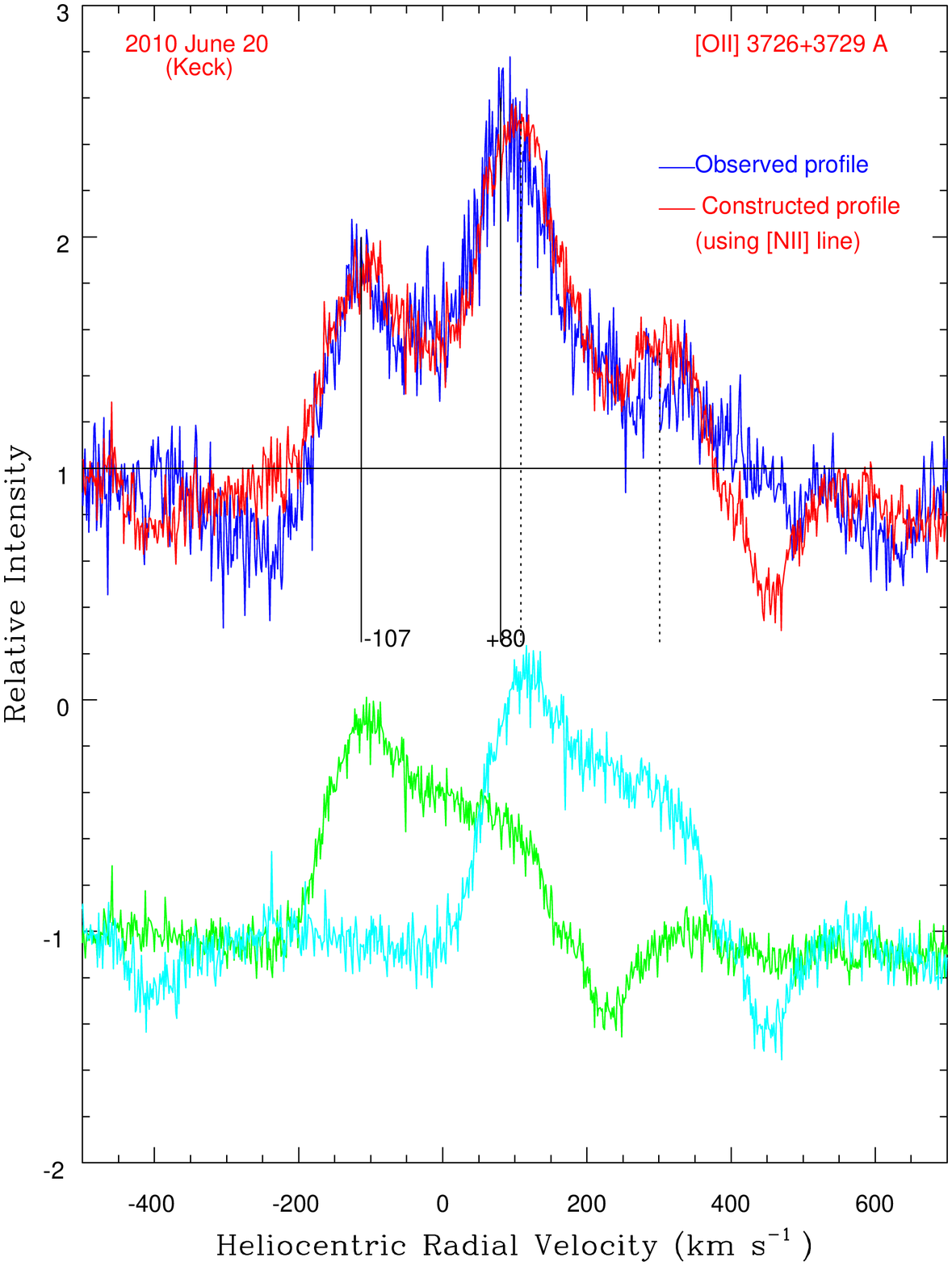}
\caption{Left panel: Low resolution spectra showing  [S II], [N II] lines on 2009 Feb 12. 
Middle panel: Keck spectrum obtained at deep minimum on 2010 Feb 23 showing 
[N II] lines (red \& cyan). The two peaks in [N II] profiles are obvious. 
The [N II] line region was amplified, and shifted up, showing variability in profile (and flux).
 The radial velocity range 
of [N II]. The arrow at 22 km/s denotes the normal stellar photospheric 
radial velocity. Right panel: the [O II] blend of 3726, 3729 and the process of estimating the line 
ratio using the [N II] $\lambda$6583 line is illustrated. 
}
   \label{fig3}
\end{center}
\end{figure}

\section{Discussion \& Questions}
Forbidden lines (low density gas) are general features of all RCBs - many unanswered questions remain. Why 
does low-density ($N_e\sim 100$ cm$^{-3}$) nebular gas has such high velocities of expansion (rotation)?. If it is 
ejected like a PN, it should have a much lower expansion velocity of $\sim$20-30 km/s (or even born-again AGB?). 
Could it have been ejected in the course of white dwarf mergers? - some merger models of CO (0.8 M$_\odot$) 
+ He (0.4 M$_\odot$) white dwarfs,  show large mass loss and formation of hot corona and a 
disk around the resultant compact object.

\noindent {Acknowledgements:} N. Kameswara Rao would like to thank Anibal Garc\'ia-Hern\'andez for help with the poster preparation. We also would like to thank Philip McQueen, Anita Cochran, and D. K. Sahu for getting the observations for us.

\end{document}